\documentclass[12pt]{article}

\usepackage{amssymb}
\usepackage[dvips]{graphicx}

\newcommand {\beq}{\begin{equation}}
\newcommand {\eeq}{\end{equation}}
\newcommand {\beqa}{\begin{eqnarray}}
\newcommand {\eeqa}{\end{eqnarray}}

\newcommand {\Tr}{\mbox{Tr}}

\newcommand {\ee}{\mbox{e}}

\newcommand {\dd}{\mbox{d}}

\newcommand {\defeq}{\stackrel{\rm def}{=}}

\newcommand {\new}{{\rm new}}
\begin{document}
\setlength{\oddsidemargin}{0cm}
\setlength{\baselineskip}{7mm}

\begin{titlepage}
 \renewcommand{\thefootnote}{\fnsymbol{footnote}}
\begin{normalsize}
\begin{flushright}
\begin{tabular}{l}
NBI-HE-01-02\\
HU-EP-01/08\\
hep-th/0104260\\
%\hfill{ }\\
April 2001
\end{tabular}
\end{flushright}
  \end{normalsize}

\vspace*{0cm}
    \begin{Large}
       \begin{center}
         {On the Spontaneous Breakdown of Lorentz Symmetry\\
in Matrix Models of Superstrings}
%\\
%\vspace*{2mm}
%         {4D SU($N$) Super Yang-Mills Theory} \\
       \end{center}
    \end{Large}
\vspace{2mm}

\begin{center}
J. A{\scshape mbj\o rn}$^{1)}$\footnote
            {e-mail address : ambjorn@nbi.dk},
K.N. A{\scshape nagnostopoulos}$^{2)}$\footnote
            {e-mail address : konstant@physics.uoc.gr},\\
W. B{\scshape ietenholz}$^{3)}$\footnote
            {e-mail address : bietenho@physik.hu-berlin.de},
F. H{\scshape ofheinz}$^{3,4)}$\footnote
            {e-mail address : hofheinz@physik.hu-berlin.de}
           {\scshape and}
           J. N{\scshape ishimura}$^{1)}$\footnote{
On leave from Department
of Physics, Nagoya University, Nagoya 464-8602, Japan,\\
e-mail address : nisimura@nbi.dk}\\
      \vspace{1cm}
      $^{1)}$ {\itshape The Niels Bohr Institute, Copenhagen University,} \\
              {\itshape Blegdamsvej 17, DK-2100 Copenhagen \O, Denmark}\\
      $^{2)}$ {\itshape Department of Physics, University of Crete,}\\
              {\itshape P.O. Box 2208, GR-71003 Heraklion, Greece}\\
      $^{3)}$ {\itshape Institut f\"{u}r Physik, Humboldt Universit\"{a}t,}\\
              {\itshape Invalidenstr. 110, D-10115 Berlin, Germany}\\
      $^{4)}$ {\itshape Fachbereich Physik, Freie Universit\"{a}t Berlin,}\\
              {\itshape Arnimallee 14, D-14195 Berlin, Germany}
\end{center}

\hspace{5cm}

%\vfill

\begin{abstract}
\noindent
In string or M theories, the spontaneous breaking of 10D or 11D
Lorentz symmetry is required to describe our space-time. A
direct approach to this issue is provided by the IIB matrix
model. We study its 4D version, which corresponds to the
zero volume limit of 4D super $SU(N)$ Yang-Mills theory. Based on
the moment of inertia as a criterion, spontaneous symmetry 
breaking (SSB) seems to occur, so that only one extended direction 
remains, as first observed by Bialas, Burda et al. 
However, using Wilson loops as probes of space-time we do not 
observe any sign of SSB in Monte Carlo simulations where $N$ 
is as large as 48. This agrees with an earlier observation that 
the phase of the fermionic integral, which is absent in the 4D 
model, should play a crucial r\^{o}le if SSB of Lorentz symmetry
really occurs in the 10D IIB matrix model.
\end{abstract}
\vfill
\end{titlepage}
\vfil\eject

\setcounter{footnote}{0}
\section{Introduction}
\setcounter{equation}{0}

\renewcommand{\thefootnote}{\arabic{footnote}} 

Matrix models \cite{BFSS,IKKT} are considered
the most promising candidate 
for a nonperturbative definition of string/M theories.
They may play an analogous r\^{o}le as 
lattice gauge theory does in quantum field theory.
One of the most fundamental questions that can be addressed
using these models is the issue of spontaneous breakdown of
Lorentz invariance, which is required to occur
in order for these theories in 10 (or 11) dimensions
to describe our four-dimensional space-time.
For early works which address this issue using
string field theory, see Refs.\ \cite{SFT}.

The IIB matrix model (or IKKT model) \cite{IKKT},
which is conjectured to
describe type IIB superstrings nonperturbatively,
is a supersymmetric matrix model
composed of 10 bosonic matrices 
and 16 fermionic matrices, 
which can be obtained formally by taking the zero-volume 
limit of 10D super SU($N$) Yang-Mills theory.
\footnote{Dimensionally reduced Yang-Mills theories
were first studied in Refs. \cite{YMd}.}
This model is particularly suitable for the
study of spontaneous symmetry breaking (SSB) of Lorentz 
invariance,\footnote{When one defines the IIB matrix model
nonperturbatively, a Wick rotation to
Euclidean signature is needed.
Hence by Lorentz symmetry we actually mean SO(10) symmetry.}
%This is also the case for other large $N$ reduced models discussed below.}
since it is mani-festly invariant under SO(10) transformations,
which transform the bosonic and fermionic matrices 
as a vector and as a Majorana-Weyl spinor, respectively.
The bosonic matrices represent the dynamically generated
space-time. A $d$-dimensional space-time is described by
configurations with $d$ bosonic matrices 
having much broader eigenvalue distributions
than the other $(10-d)$ matrices, up to some SO(10) 
transformation.\footnote{Throughout this paper, 
we denote the initial dimensionality
by $D$ and the dimension after a possible SSB as $d$.}
Our four-dimensional space-time
may be accounted for, if the $d=4$ configurations
(in the above sense) 
dominate the integration over the bosonic matrices.
It was found recently that the IIB matrix model
is indeed endowed with a natural mechanism that may realize
such a scenario \cite{NV,brane}.
%Here our four-dimensional space-time is realized as a `brane'
%in ten-dimensional space-time, and therefore

The realization of our space-time as a `brane' in a 
higher-dimensional space-time has attracted much attention
as an alternative to the more conventional approach in string theory
using compactification (see Ref.\ \cite{RS} and references therein).
It turned out that such a set-up has many phenomenological
advantages, including possible mechanisms 
which may solve the cosmological problem and the hierarchy problem.
However, the dynamical origin of the brane has not been discussed so far.
The IIB matrix model enables us to investigate whether
a four-dimensional space-time emerges
{\em dynamically} as a brane 
in ten-dimensional 
type IIB superstring theory through some nonperturbative effects
\footnote{A possible obstacle may be
that gravitons propagate in ten-dimensional space-time, 
and hence one fails to reproduce the observed four-dimensional Newton's law.
In Ref.\ \cite{IIKK}, it was demonstrated that this obstacle
can be avoided in the case of D3-brane backgrounds
due to the mechanism of Ref.\ \cite{RS}.}.

The spontaneous breakdown of Lorentz symmetry in matrix models
has been addressed first in the bosonic case \cite{HNT},
where fermionic matrices are omitted
(for recent work on the bosonic model, see Ref.\ \cite{bos,AW}).
There the absence of SSB has been established  
by both an analytical method (to all orders in a $1/D$ expansion)
and by Monte Carlo simulations.
The same numerical result was obtained in the 6D and 10D SUSY
matrix models \cite{branched}, albeit with some 
simplifications to enable simulations at large $N$.
The fermion integrals are complex in general in these cases,
and the simulations were carried out including only the modulus,
but omitting the phase. In addition, a low-energy effective theory 
was used in order to further reduce the computational effort.
%Although we have a certain numerical evidence that
%the one-loop approximation captures the low-energy dynamics
%of the supersymmetric matrix models \cite{AABHN},
%cumulative effects of higher loop corrections may eventually lead 
%to a non-trivial phenomenon such as SSB of Lorentz symmetry.

In Ref.\ \cite{AABHN} we presented Monte Carlo simulations 
of the 4D version of the IIB matrix model,
which is a supersymmetric matrix model obtained
from the zero-volume limit of 4D super SU($N$) Yang-Mills theory.
We were able to study the model with $N=16,24,32,48$
without any simplifications.
%by carefully implementing the Hybrid R algorithm.
These values of $N$ turned out to be sufficiently large to
extract the large $N$ behavior of the space-time structure 
and to reveal the large $N$ scaling for a number
of Wilson loop correlators.

Recently, it has been reported for the 4D SUSY model up to $N=8$
that the space-time is observed to be one-dimensional,
if one selects configurations with large extent
from the ensemble \cite{Burda:2000mn}.
%On the other hand, if one considers configurations with
%a typical extent, the space-time is observed to be 
%four-dimensional, as we are going to show (see in particular 
%Figure \ref{Tnew-EV-N}).
In the $D$-dimensional SUSY models in general, $D=4,6,10$,
configurations with large extent are suppressed only by the
power $-(2D-5)$, independent of $N$ \cite{Eigen}.
Therefore, the observed anisotropic configurations
may play some r\^{o}le in the large $N$ limit,
and such effects may also be relevant in other SUSY models,
including the IIB matrix model.

In this paper, we reconsider the issue of SSB of 
Lorentz invariance in the 4D SUSY model.
If we adopt the conventional criterion
based on the moment of inertia tensor, then
the space-time appears one-dimensional,
as suggested by the observation in Ref.~\cite{Burda:2000mn}.
This would mean that the SSB does occur at large $N$.
However, this conclusion depends on the definition 
of the order parameter, as we shall see.
%changes drastically if one makes a
%modification in the definition of the order parameter for the SSB.
%A similar problem may occur generally in
%SUSY large $N$ models including the IIB matrix model.
Thus we have to address the question which criterion for the SSB 
of Lorentz symmetry is actually physical.

%The first question is of course whether
%any signal of SSB can be observed in physical observables.
%In this regard, 
We recall that in the interpretation of 
the IIB matrix model as a string theory, 
the Wilson loops are identified 
with the string creation operators \cite{FKKT}.
Physical observables --- such as scattering amplitudes ---
should be extracted from correlation functions of Wilson loops,
which were observed to have well-defined large $N$ limits in $D=4$
\cite{AABHN}.
%and in $D=10$ \cite{KU}.
We therefore propose a ``physical criterion'' of 
SSB using Wilson loops as a probe,
and we study it by Monte Carlo simulations of the full 4D $SU(N)$ SUSY model.
As a result, we find no trend of SSB up to $N=48$.

\section{The 4D IIB matrix model}
\label{model}
\setcounter{equation}{0}

The model we investigate is a supersymmetric matrix model
obtained from the zero-volume limit of 4D
SU($N$) super Yang-Mills theory.
Its partition function is given by
\begin{eqnarray}
Z &=& \int \dd A ~  \ee ^{-S_b} \int \dd \psi \dd \bar{\psi}
~ \ee ^{- S_f } \ , \nonumber \\
S_b &=& -\frac{1}{4 g^2} \, \Tr \, [A_{\mu},A_{\nu}]^{2} \ , 
\nonumber \\
S_f  &=& - \frac{1}{g^2} \,
\Tr \, \Big( \bar{\psi}_\alpha  (\Gamma^{\mu})_{\alpha\beta} 
[A_{\mu},\psi _\beta] \Big) \ ,
\label{action}
\end{eqnarray}
where $A_\mu$ ($\mu=1 , \dots , 4$) are bosonic traceless
$N \times N$ Hermitian matrices,
and $\psi_\alpha$, $\bar{\psi}_\alpha$ ($\alpha = 1,2$) 
are fermionic traceless $N \times N$ complex matrices.
The 2 $\times$ 2 unitary matrices $\Gamma_\mu$ are 
gamma matrices after Weyl projection;
they can be given for example by
\begin{equation}
\Gamma_1 = i  \sigma_1 , \quad
\Gamma_2 = i \sigma_2 , \quad
\Gamma_3 = i  \sigma_3 , \quad
\Gamma_4 =   {\bf 1}  \ .
\label{Gamma}
\end{equation}
This model is invariant under 
4D Lorentz transformations,
where $A_{\mu}$ transforms as a vector and
$\psi _ \alpha$ as a Weyl spinor.
The model is manifestly supersymmetric,
and it also has a SU($N$) symmetry
\beq
\label{SU_N}
A_\mu  \rightarrow  V A_\mu V^\dag  ~~~;~~~
\psi _\alpha \rightarrow V \psi _\alpha V^\dag   ~~~,~~~
\bar{\psi} _\alpha \rightarrow V \bar{\psi} _\alpha V^\dag   ,
\eeq
where $V\in \mbox{SU}(N)$.
All these symmetries are inherited from the super Yang-Mills theory
before the zero-volume limit.
The model can be regarded as the four-dimensional counterpart of
the IIB matrix model.

The model is well-defined for arbitrary $N\ge 2$ without any cutoff.
This was first conjectured based on numerical results at small $N$ 
\cite{KNS}, confirmed further at larger $N$ \cite{AABHN} and 
and finally proved by Ref.\ \cite{AW}.
Therefore, the parameter $g$ --- which is the only parameter of the 
model --- can be absorbed by rescaling the variables,
\beq
\label{rescaleA}
A_\mu = g^{1/2} X _\mu  ~~~~~;~~~~~
\psi_\alpha = g^{3/4} \Psi _\alpha  \ .
\eeq
Therefore, $g$ is a scale parameter rather than
a coupling constant, i.e.\
the $g$ dependence of physical quantities is completely 
determined on dimensional grounds. The parameter
$g$ should be tuned appropriately as one sends $N$ to infinity,
so that each correlation function of Wilson loops
has a finite large $N$ limit.
This issue has been studied numerically in Ref.\ \cite{AABHN}. 
The conclusion is that the product $g ^2 N$ has to be kept constant
when taking the large $N$ limit. The tuning of $g$ was also discussed
in terms of analytical arguments
in Refs.~\cite{FKKT,AIKKT,Aoki:1999bq,KU}.

The integration over fermionic variables can be done explicitly
and the result is given by \ $\det {\cal M} $,
${\cal M}$ being a
$2(N^2-1)$ $\times$ $2(N^2-1)$ complex matrix which depends
on $A_\mu$. The matrix ${\cal M}$ is given by
\beq
{\cal M}_{a \alpha ,  b \beta} \equiv
(\Gamma _\mu) _{\alpha \beta} \, \Tr \left( t^a \,  
[ A_\mu,t^b] \right) \ ,
\label{matrix_M} 
\eeq
where $t^a$ are generators of SU($N$), and we consider
$(a\alpha)$ resp.\ $(b\beta)$ as one index.
Hence the system we want to simulate can be written
in terms of bosonic variables,
\beq
Z =  \int \dd A  ~ \ee ^{-S_b} \det {\cal M} \ .
\eeq
A crucial point is that the determinant \ $\det {\cal M}$ \ is
{\em real positive} \cite{AABHN}. This property was demonstrated
in Ref.\ \cite{AABHN}, and it had been suspected earlier \cite{KNS}.
(It also holds in other 4D SUSY models, see Ref.\ \cite{AV}
and second Ref.\ in \cite{Burda:2000mn}.)
Due to this property,
we can simulate the model using a standard algorithm for 
dealing with dynamical fermions (the so-called
Hybrid R algorithm \cite{hybridR}). In the framework of this 
algorithm, each update of a configuration is done by
solving a Hamiltonian equation for a fixed ``time'' $\tau$.
This algorithm is plagued by a systematic error due to the 
discretization of $\tau$ that we used to solve the equation numerically.
We performed simulations at three different values of 
the ``time step'' $\Delta \tau$ and we extrapolate to $\Delta \tau = 0$.
%(See Ref.\ \cite{AABHN} for more details.)

\section{SSB of Lorentz symmetry ?}
\label{ssb}
\setcounter{equation}{0}

In the IIB matrix model, the eigenvalues of the bosonic matrices
$A_\mu$ are interpreted 
as the space-time coordinates \cite{IKKT,AIKKT,Iso:2000xs}.
We adopt this point of view in the 4D model as well.
Since the matrices
$A_\mu$ are not simultaneously diagonalizable in general,
the space-time is not classical.
To extract the space-time structure we first define the 
space-time uncertainty $\Delta$ by \cite{AABHN}
\beq
\Delta ^2  = \frac{1}{N} \, \Tr (A_\mu^{~2})
- \max_{U \in \mbox{\scriptsize SU}(N)} \frac{1}{N}
\sum_i \{ (U A_\mu U^\dagger)_{ii}  \} ^2 \ ,
\label{maxU}
\eeq
which is invariant under Lorentz transformations and under the
SU($N$) transformations (\ref{SU_N}).
Formula (\ref{maxU}) has been derived in Ref.\ \cite{HNT}
based on the analogy to quantum mechanics,
considering $A_\mu$ as an operator acting on a space of states.
As a natural property,
$\Delta ^2$ vanishes if and only if the matrices $A_\mu$ are 
diagonalizable simultaneously.
For each configuration $A_\mu$ generated by a Monte Carlo simulation, 
we maximize 
$\sum_i \{ (U A_\mu U^\dagger)_{ii}  \} ^2 $ with respect to 
the SU($N$) matrix $U$.
We denote the matrix which yields the maximum as $U_{\max}$,
and we define 
\beq
x_{i \mu} = (U_{\max} A_\mu U_{\max}^\dagger)_{ii}
\eeq
as the space-time coordinates of $N$ points $x_{i}$
($i=1,\cdots , N$) in four-dimensional space-time.

\begin{figure}[hbt]
%\vspace{6mm}
  \begin{center}
    \includegraphics[width=.41\linewidth]{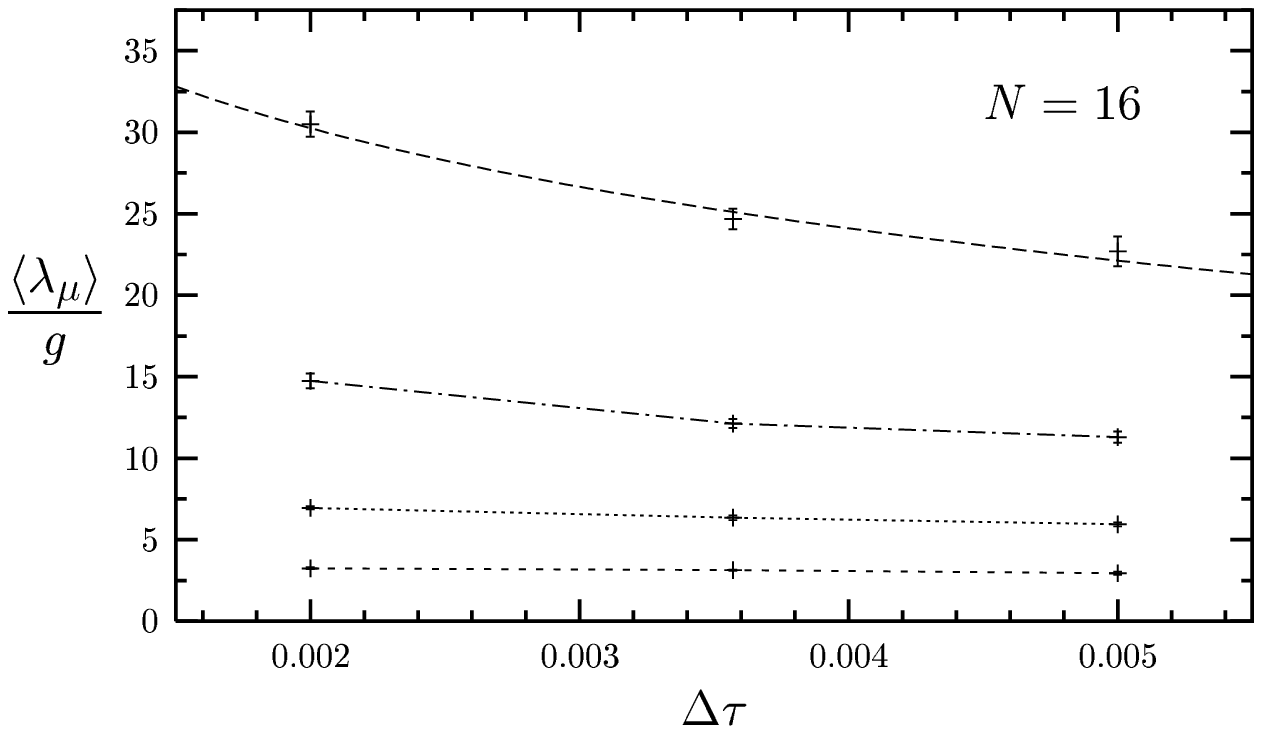} \\
\vspace{6mm}
    \includegraphics[width=.41\linewidth]{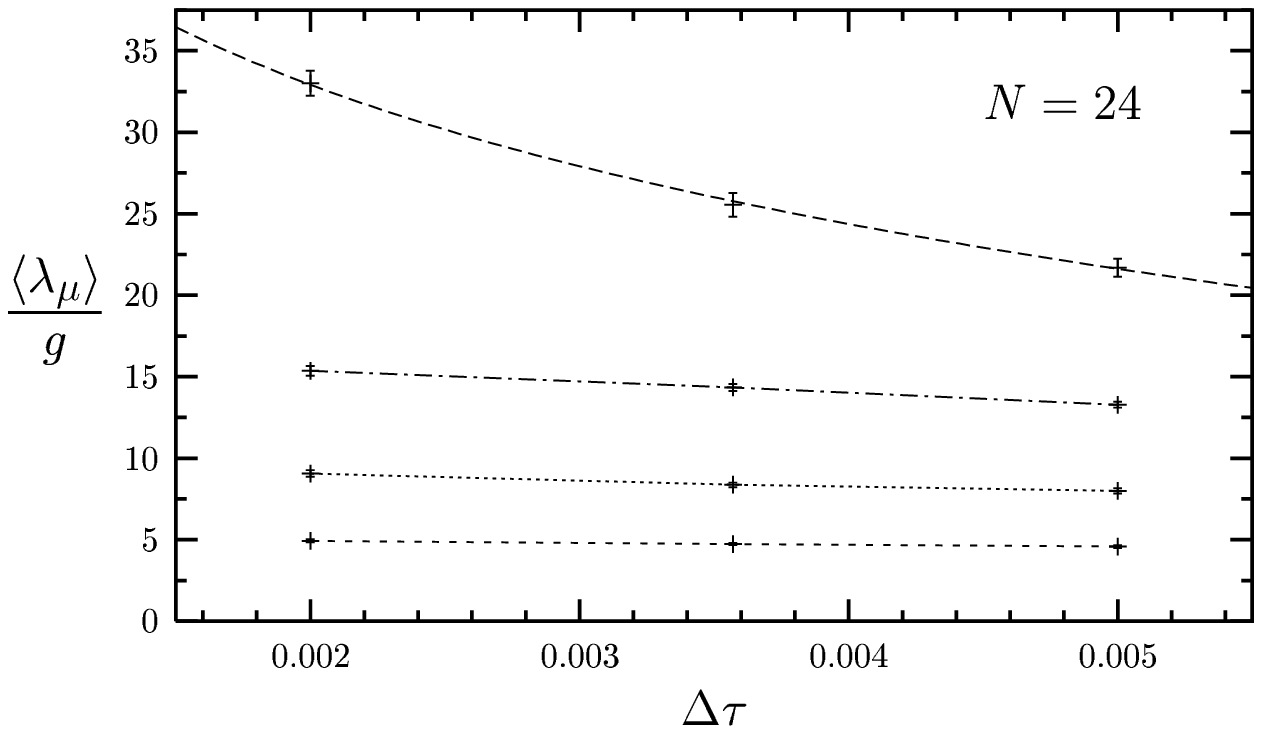} \\
\vspace{6mm}
    \includegraphics[width=.41\linewidth]{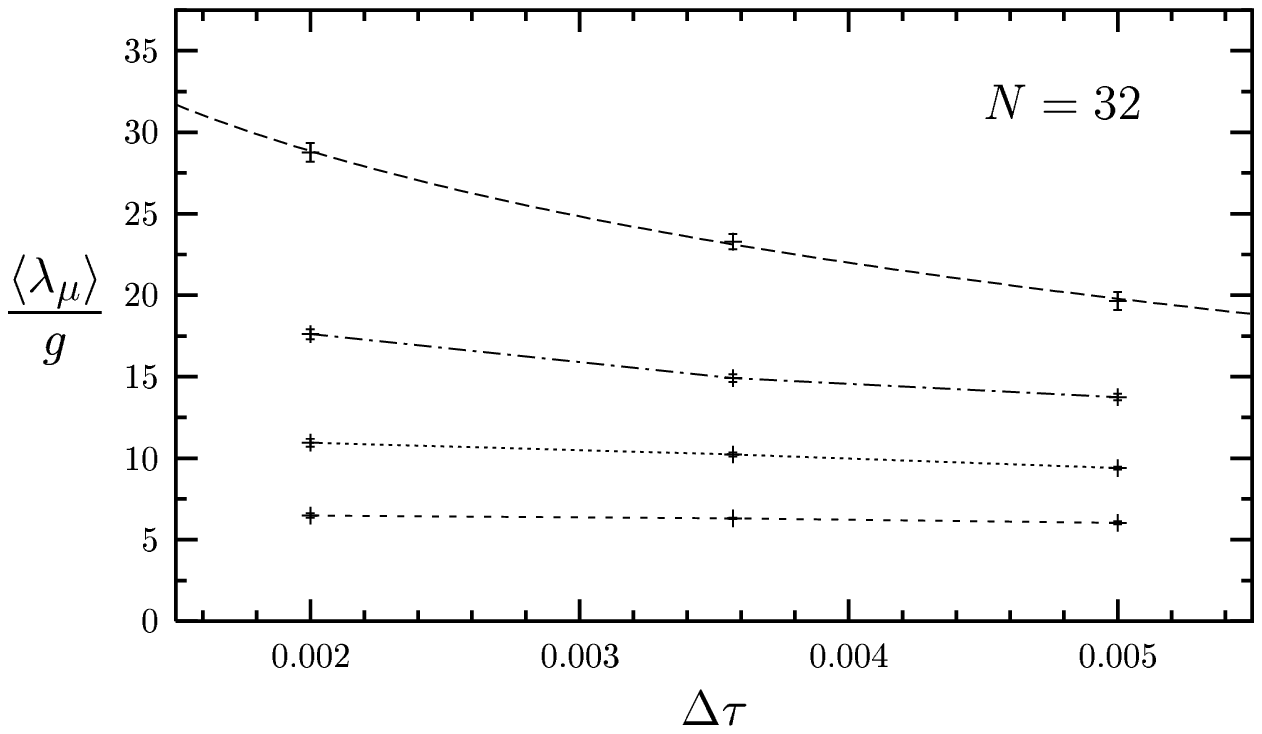}
  \end{center}
%\hspace{2cm}
%\def\fpsangle{0}
%\epsfxsize=80mm
%\fpsbox{t-N16.eps} \\
%\vspace{5mm}
%\hspace{2cm}
%\def\fpsangle{0}
%\epsfxsize=80mm
%\fpsbox{t-N24.eps} \\
%\vspace{3mm}
%\hspace{2cm}
%\def\fpsangle{0}
%\epsfxsize=80mm
%\fpsbox{t-N32.eps}
\vspace{-5mm}
\caption{\it{The eigenvalues of the tensor of inertia $T$
as functions of the algorithmic ``time step'' $\Delta \tau$
at $N=16$, $24$ and $32$. 
The dashed lines on top represent the fits to eq.\ (\ref{divlambda}),
confirming that the largest eigenvalue diverges logarithmically
in the limit $\Delta \tau \rightarrow 0$.
(The other lines are drawn to guide the eye.)}}
\label{T-EV}
\end{figure}

In order to search the spontaneous breakdown of Lorentz symmetry,
we first consider the moment of inertia tensor 
of $N$ points $x_{i}$. % ($i=1,\cdots ,N$).
It can be defined as
\beq
T _{\mu\nu} = \frac{2}{N(N-1)}  \sum _{i < j} 
(x_{i\mu} - x_{j\mu} ) (x_{i\nu} - x_{j\nu} ) \ ,
\label{defT}
\eeq
which is a $D \times D$ real symmetric matrix.\footnote{This quantity has 
also been studied in Refs.\ \cite{Aoki:1999bq,branched}.
Alternatively, 
one may define the moment of inertia tensor 
by $I_{\mu\nu}=\frac{1}{N}\Tr (A_\mu A_\nu)$, as it was done in Refs.\
\cite{HNT,Burda:2000mn}.
We have also measured the eigenvalues of that tensor, and the spectra 
are in qualitative agreement with the eigenvalues that we present
here (based on definition (\ref{defT})).\label{AmuAnu}}
The $D$ eigenvalues $\lambda _1 > \lambda _2 > \dots > \lambda _{D} >  0$ 
of the tensor $T$ represent the principal moments of inertia.
%We measure $\lambda _\mu$ for each configuration and 
We take the average $\langle \lambda _\mu \rangle$
over all configurations generated
by the Monte Carlo simulation.

In Figure \ref{T-EV} we plot the results for 
$\langle \lambda_{\mu} \rangle / g$
against $\Delta \tau$ at $N=16$, 24 and 32.
(In all the Figures in this Section, we plot dimensionless quantities,
so that they do not depend on the choice of the scale parameter $g$.)
As $\Delta \tau$ vanishes, $\langle \lambda _1 \rangle / g$
is observed to diverge as 
\beq
\langle \lambda _1 \rangle /g \sim  - c_1 \ln \Delta \tau + c_0
\label{divlambda}
\eeq
(where $c_0$, $c_1$ are constant in $\Delta \tau$),
while the other eigenvalues converge.\footnote{We note 
that based on our data, a logarithmic
divergence of $\langle \lambda_{2}\rangle$ 
is unlikely, but it cannot
be absolutely excluded. This slight uncertainty is still a little
stronger if we use the definition 
$I_{\mu\nu}=\frac{1}{N} \Tr (A_{\mu} A_{\nu})$
for the moment of inertia tensor.}
We recall that the extent of the space-time $R$ defined by
\beq
R^2  = \frac{2}{N(N-1)} 
\left\langle \sum_{i < j}  (x_i - x_j)^2   \right\rangle 
\label{Rdef}
\eeq
was found to be divergent \cite{AABHN}.
Since $R^2 = \sum_\mu \langle \lambda _\mu \rangle$,
at least $\langle \lambda _1 \rangle$ 
has to diverge as $\Delta \tau \rightarrow 0$.
What we observe here is that $\langle \lambda_{1}\rangle$ 
is indeed the only 
divergent eigenvalue.
This is consistent with the observation in Ref.\ \cite{Burda:2000mn}.

\begin{figure}[hbt]
  \begin{center}
    \includegraphics[width=.45\linewidth]{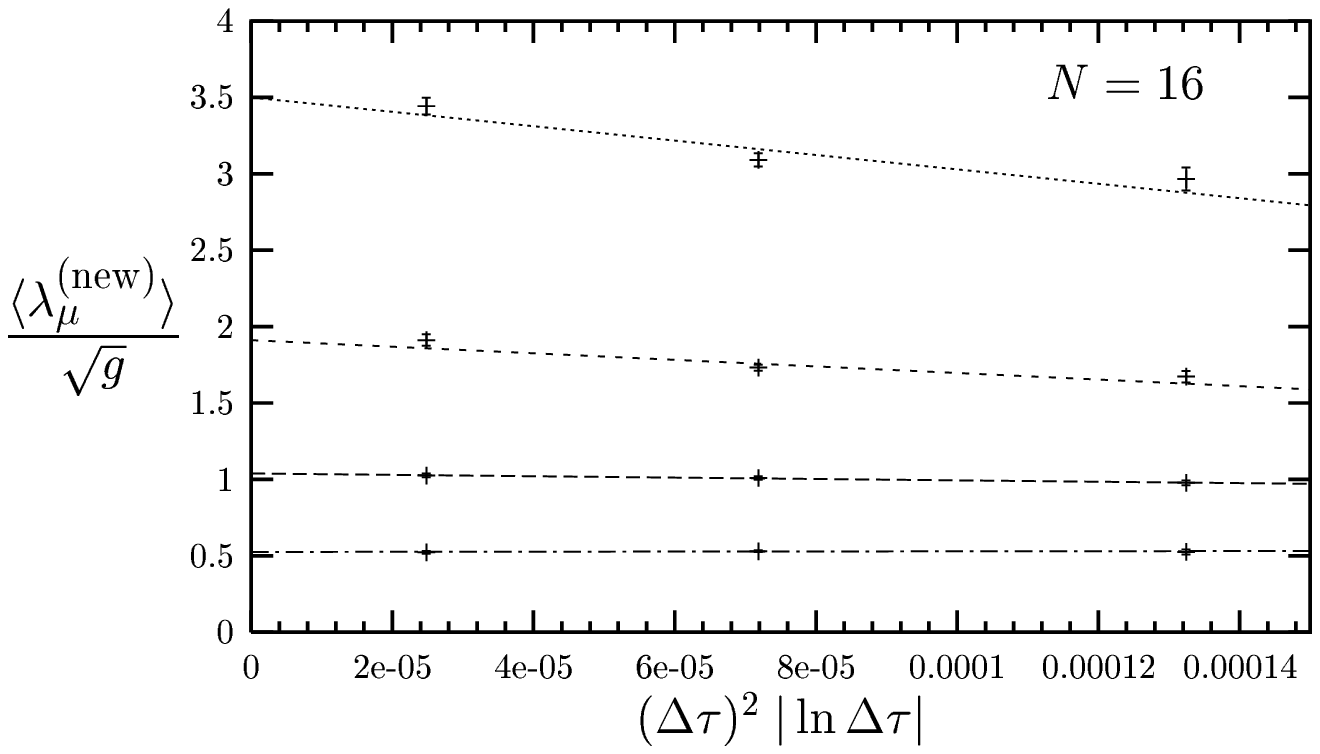} \\
\vspace{2mm}
    \includegraphics[width=.45\linewidth]{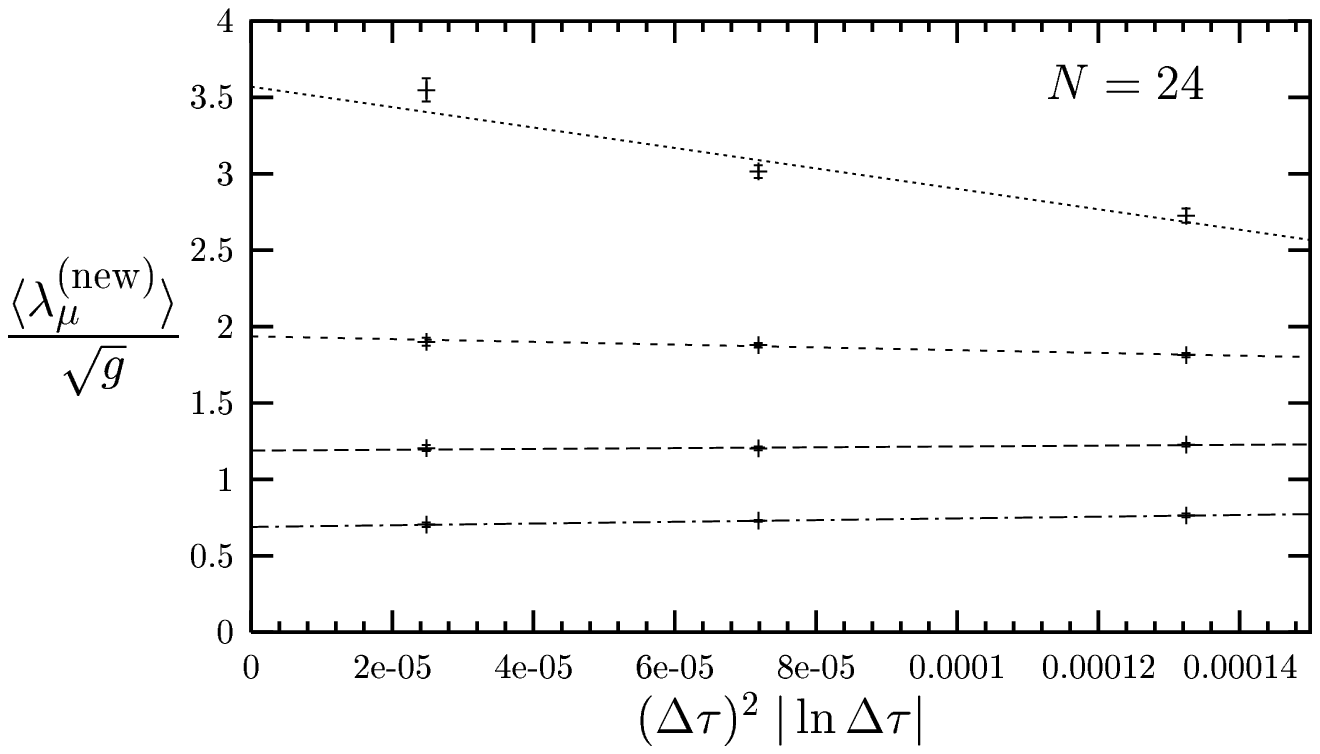} \\
\vspace{2mm}
    \includegraphics[width=.45\linewidth]{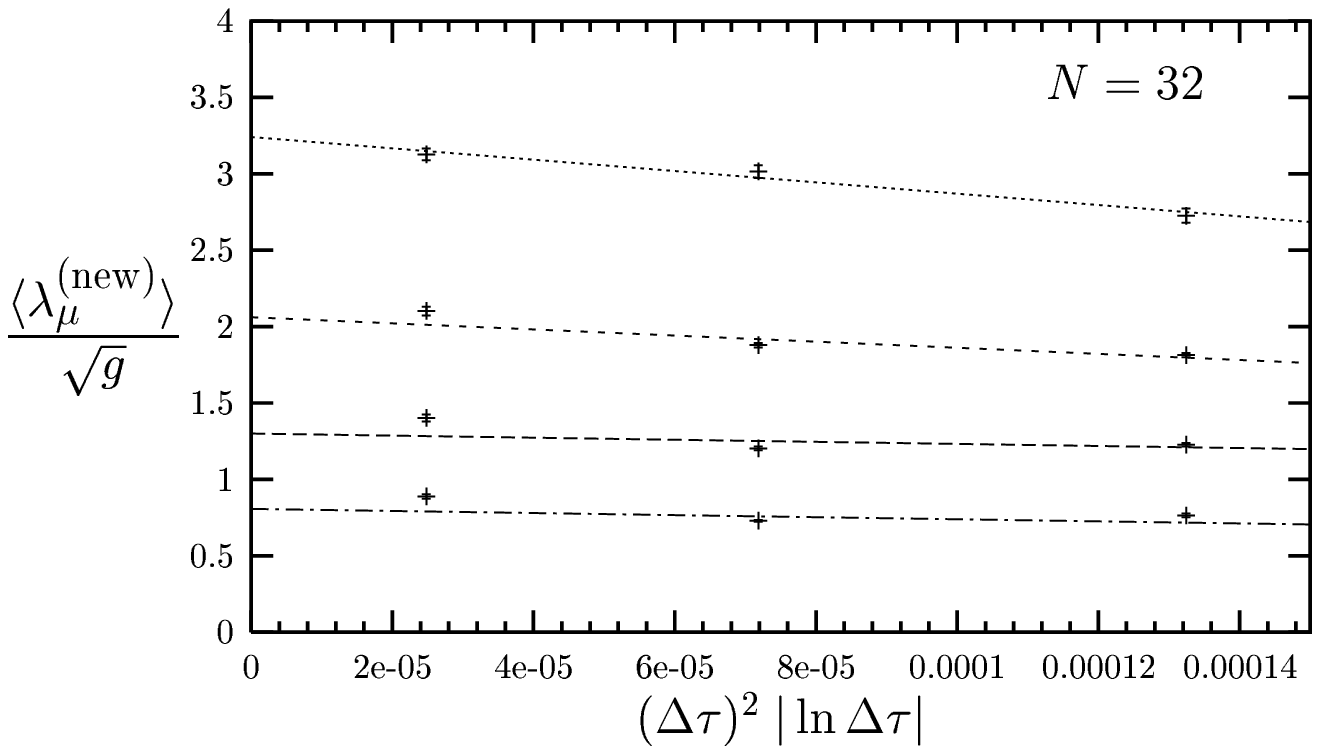}
  \end{center}
\vspace{-6mm}
\caption{\it{The eigenvalues of the new tensor
$T^{({\rm new})}_{\mu, \nu}$ as functions of the algorithmic 
``time step'' $\Delta \tau$ at $N=16$, $24$ and $32$. 
The straight lines are the fits to eq.\ (\ref{extrapol_assumption}).
We see that now all eigenvalues converge
in the limit $\Delta \tau \rightarrow 0$.}}
%!!!the y axis should be labeled by 
%$\langle \lambda_\mu ^{(new)} \rangle /\sqrt{g}$ !!!}}
\label{Tnew-EV}
\end{figure}

Let us introduce the probability distribution for the distance of 
two space-time points as
\beq
\rho (r) = \frac{2}{N(N-1)} \left\langle \sum_{i < j}  
\delta \left(r - \sqrt{(x_i - x_j)^2}\right)  \right\rangle \ .
\label{rho_r}
\eeq
Then $R^2$ can be written as
\beq
R^2 =  \int_0 ^\infty  \dd r ~ r^2 \rho (r) \ .
\eeq
The observed logarithmic divergence of $R^2$ is consistent with
the asymptotic behavior 
\beq
\rho (r) \sim r ^{-3} \ , 
\label{asym_rho}
\eeq
which was predicted
analytically \cite{Eigen}. 
Based on this observation, a modified definition for the extent of 
the space-time has been introduced in Ref.\ \cite{AABHN},
\beq
R _{\new} = \frac{2}{N(N-1)} 
\left\langle \sum_{i < j}  \sqrt{(x_i - x_j)^2}   \right\rangle 
= \int_0 ^\infty  \dd r ~ r \rho (r) \ ,
\label{Rnewdef}
\eeq
which turned out to be finite --- as expected from 
relation (\ref{asym_rho}).
The large $N$ behavior of this quantity has been
observed to amount to $R_{\new}/\sqrt{g} = 3.30(1) \cdot N^{1/4}$,
which is consistent with the prediction based on 
the low-energy effective theory \cite{AIKKT}.

\begin{figure}[hbt]
  \begin{center}
    \includegraphics[width=.55\linewidth]{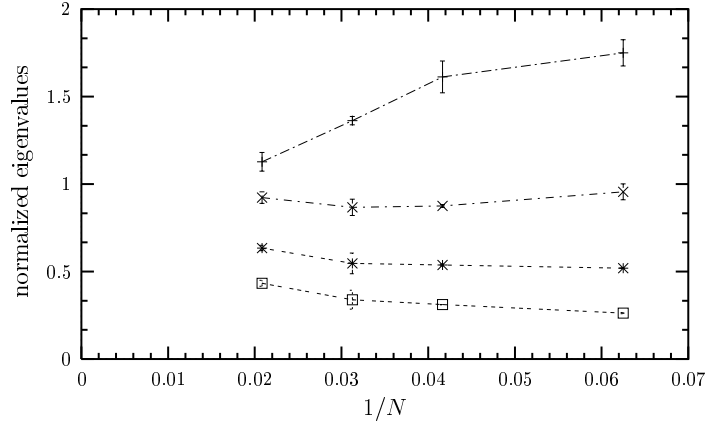}
  \end{center}
\vspace{-6mm}
\caption{\it{The eigenvalues of the new tensor
$T^{({\rm new})}_{\mu ,\nu}$, extrapolated to 
$\Delta \tau =0$ and normalized by the factor 
$(\sqrt{g} N^{1/4})^{-1}$. %depending on $N$. 
We also include the results for $N=48$, which are obtained
at $\Delta \tau = 0.002$ (our statistics at other
values of $\Delta \tau$ is insufficient for a sensible
extrapolation in this case).}}
%The lines are drawn to guide the eye.}}
\label{Tnew-EV-N}
\end{figure}

This motivates us to define analogously a new tensor
\beq
T ^{(\new)}_{\mu\nu} = \frac{2}{N(N-1)} 
\left\langle \sum_{i < j}  
\frac{(x_{i\mu} - x_{j\mu} ) (x_{i\nu} - x_{j\nu} )}
{\sqrt{(x_i - x_j)^2}}   \right\rangle \ .
\label{Tnewdef}
\eeq
Let us denote the $D$ eigenvalues of the tensor $T^{(\new)}$
as $\lambda^{(\new)}_1 > \lambda^{(\new)}_2 > \cdots 
> \lambda^{(\new)}_{D} >  0$. Due to the relation $\sum_{\mu} 
\lambda^{(\new)}_\mu = R_{\new}$, all the eigenvalues are expected to 
converge. In Figure \ref{Tnew-EV} we show the results for 
$\langle \lambda^{(\new)}_\mu \rangle / \sqrt{g}$ \
% against $\Delta \tau$, 
again at $N=16$, 24 and 32.
%This Figure confirms that all eigenvalues do converge.
We carried out an extrapolation to 
$\Delta \tau =0$ assuming the observable ${\cal O}(\Delta \tau)$ 
at small $\Delta \tau$ to behave as \cite{AABHN}
\beq
{\cal O}(\Delta \tau) - {\cal O}(\Delta \tau=0) 
%\sim (\Delta \tau)^2 \cdot
%\langle \tr (A_\mu^{~2}) \rangle _{\Delta \tau}
\propto (\Delta \tau)^2 \cdot |\ln \Delta \tau | \ .
\label{extrapol_assumption}
\eeq
%at small $\Delta \tau$.
In Figure \ref{Tnew-EV-N} we plot the extrapolated and 
normalized eigenvalues
$\langle \lambda^{(\new)}_\mu \rangle / (\sqrt{g} N^{1/4})$ 
against $1/N$.
(This is the normalization needed for a finite large $N$ limit.)
We observe that they move closer together as $N$ increases.
Therefore we cannot recognize any trend for SSB.

\section{A physical criterion in terms of Wilson loops}
\label{physical}
\setcounter{equation}{0}

The results in the previous Section reveal a subtlety
in the issue of SSB in the 4D SUSY model.
The crucial question is whether any signal of SSB can be
probed by physical quantities, such as scattering amplitudes.
Therefore we have to reconsider how the IIB
matrix model is interpreted as a string theory.
In Ref.\ \cite{FKKT} it has been demonstrated that
Wilson loops in matrix models can be identified with
string creation operators in string theory.
Hence Wilson loop correlation functions are the only objects
with a direct physical interpretation in string theory.
So we should ask whether any signal of SSB can be probed
by Wilson loops.

\begin{figure}[hbt]
%\vspace{-3mm}
  \begin{center}
    \includegraphics[width=.46\linewidth]{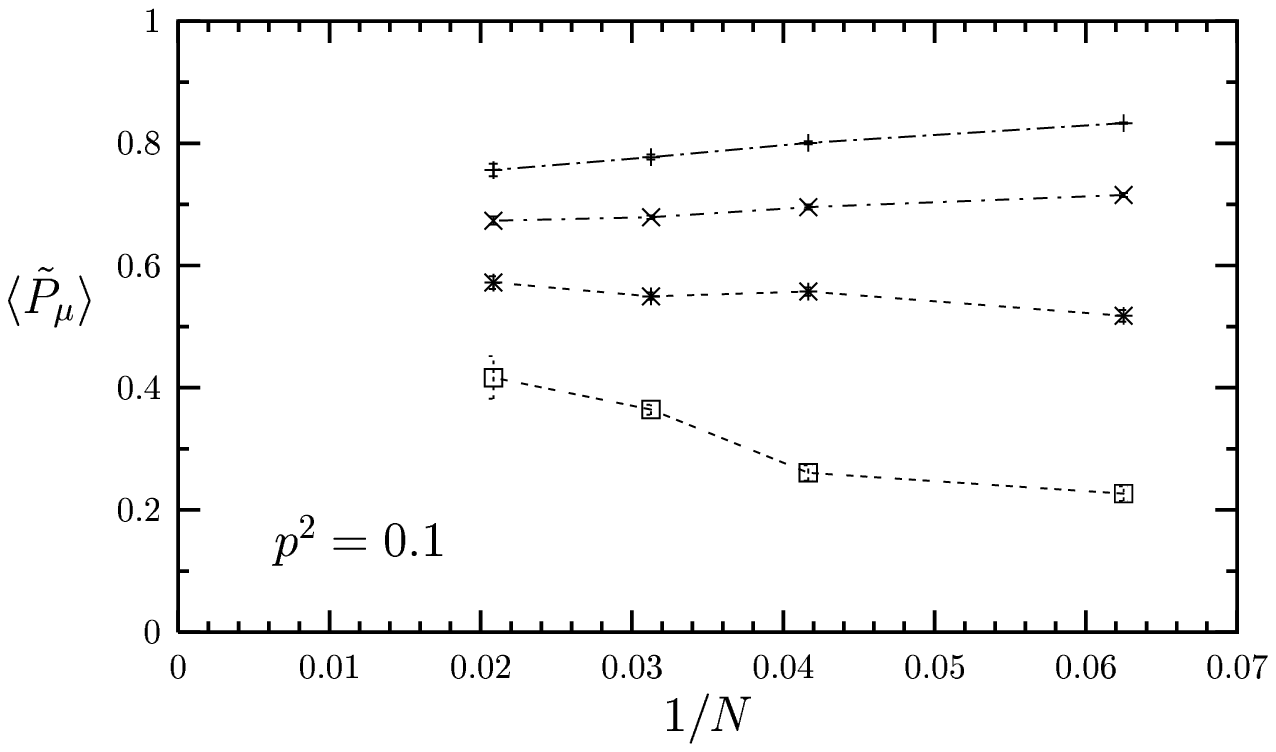} \\
\vspace{2mm}
    \includegraphics[width=.46\linewidth]{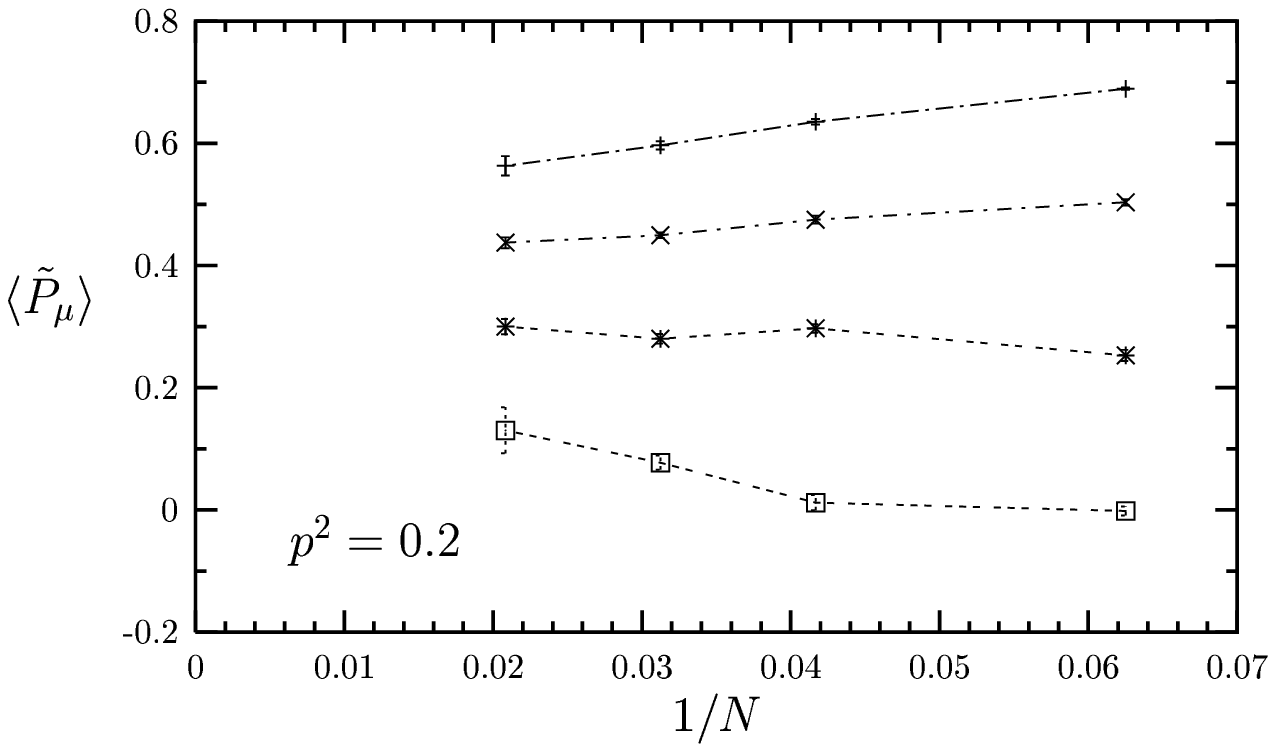} \\
\vspace{5mm}
    \includegraphics[width=.46\linewidth]{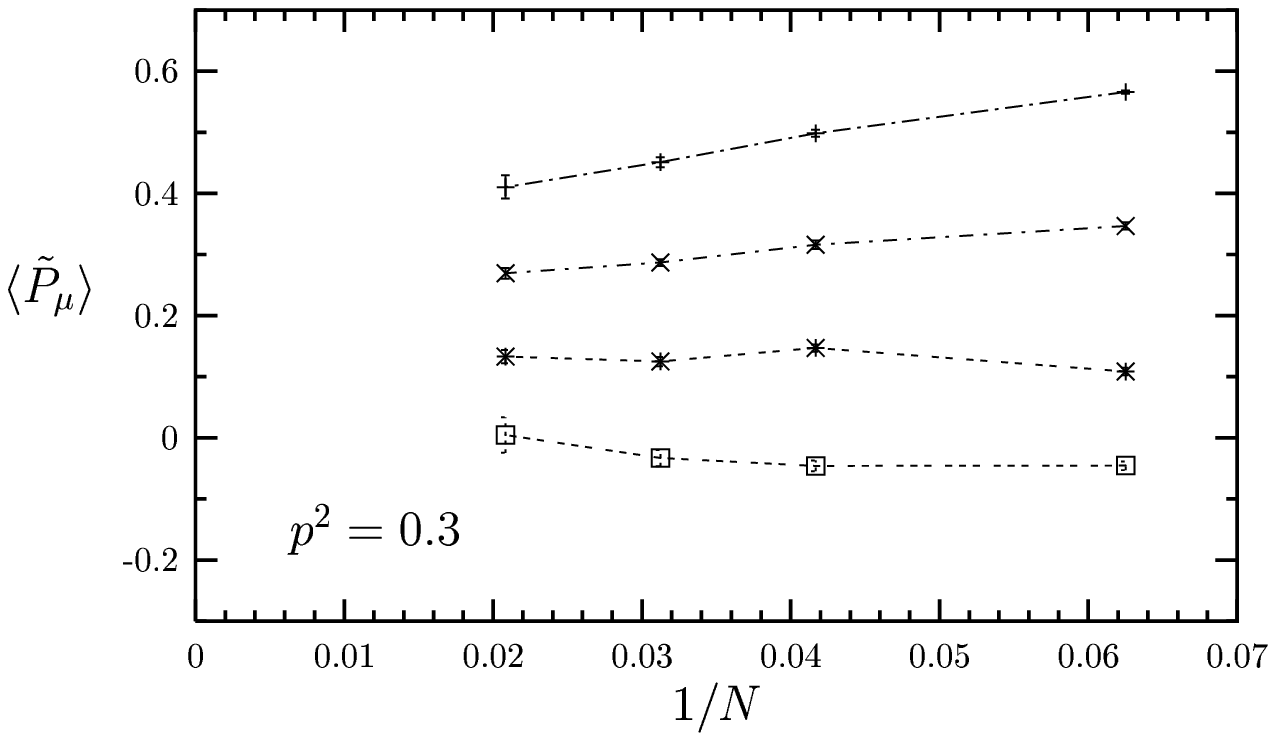} 
  \end{center}
\vspace{-6mm}
\caption{\it{The four Polyakov lines $\langle \tilde P_{\mu} (p) \rangle$
($\mu = 1 , \dots , 4$) at $p = 0.316$, $0.447$, $0.548$,
where we take the scaling parameter $g$ to be $g = (N/48)^{-1/2}$. 
The Polyakov lines become approximately equidistant
and they move closer together as $N$ increases, hence we do not
see any signal for SSB of Lorentz symmetry.
(The lines are drawn to guide the eye.)}}
\label{P-k-fig}
\end{figure}

\begin{figure}[hbt]
  \begin{center}
    \includegraphics[width=.46\linewidth]{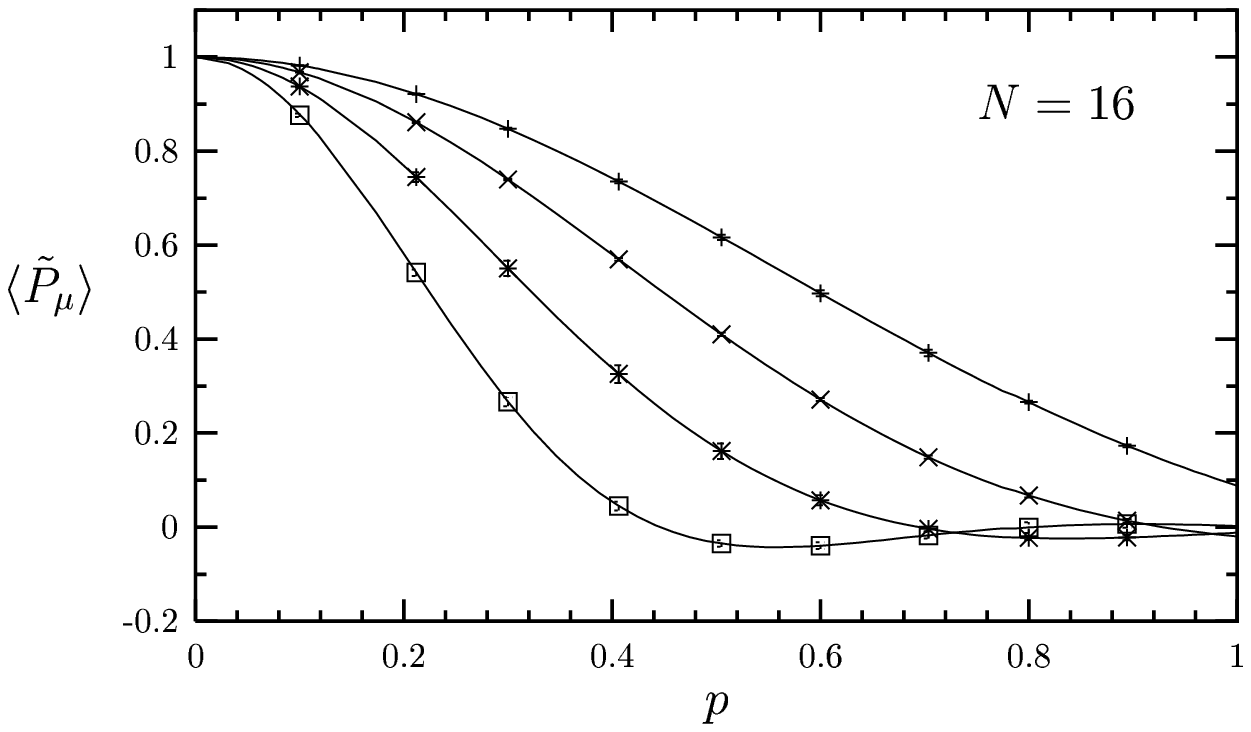} \\
\vspace{2mm}
    \includegraphics[width=.46\linewidth]{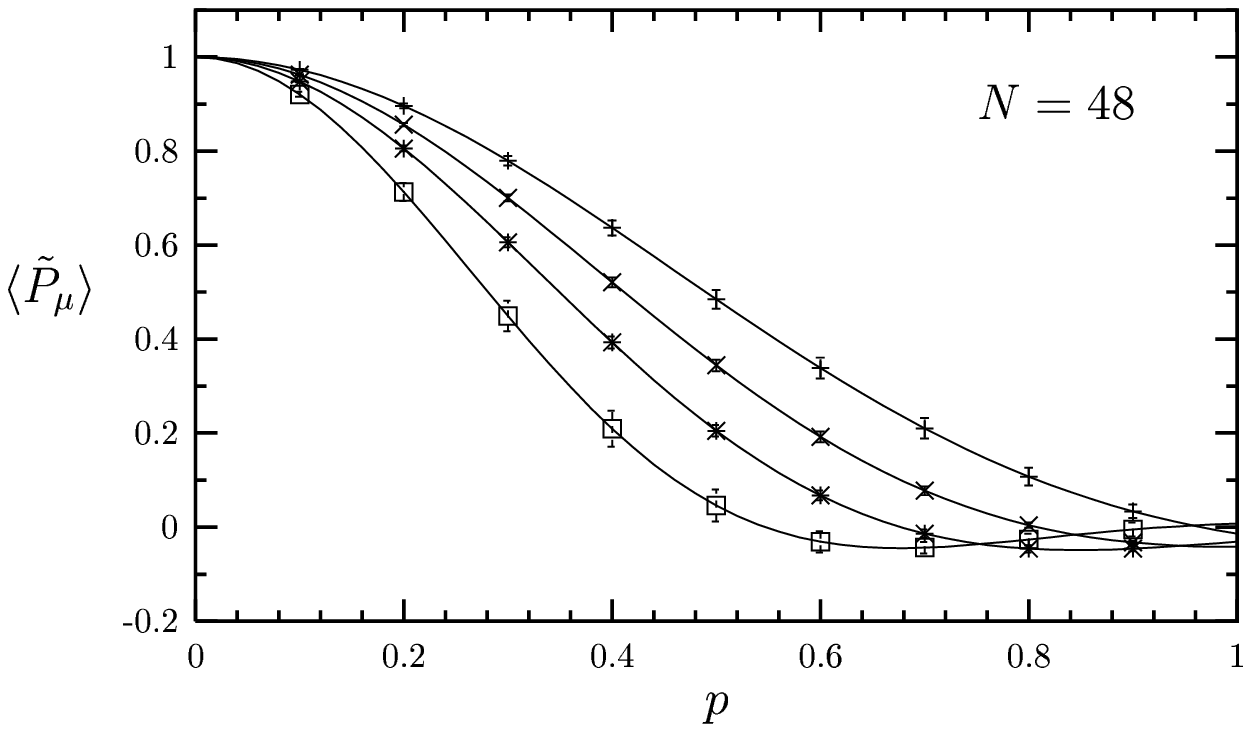}
  \end{center}
\vspace{-6mm}
\caption{\it{The Polyakov lines $\langle \tilde 
P_{\mu} (p) \rangle$ ($\mu = 1,  \dots , 4$)
at $N=16$ and at $N=48$. 
They move closer together as $N$
increases, in agreement with the proposed scenario that they 
coincide in the large $N$ limit. (The curves were measured
quasi-continuously, and we show a few error bars as 
examples.)}}
\label{P-k-fig2}
\end{figure}

We recall that the extent of the space-time can be probed by
the vacuum expectation value (VEV) of the ``Polyakov line'' \cite{AABHN}
\beq
P (\vec{p})
= \frac{1}{N}  \Tr \, \exp (i \, p_\mu A_\mu) \ ,
\label{Pol1}
\eeq
where $p_\mu$ represents 
the total dimensionful (and hence ``physical'') momentum
carried by the string.\footnote{This momentum $p$ is identical 
to $k_{\rm phys}(=k/\sqrt{g})$ introduced in Ref.\ 
\cite{AABHN}. We take
this opportunity to correct a typo in Fig.\ 5 of Ref.\ \cite{AABHN}.
The label for the horizontal axis should be $k / \sqrt{g}$ instead
of $k ^2  / g$.
This typo has propagated to our review 
articles \cite{Ambjorn:2001xj,Ambjorn:2000dj} as well.
Fig.\ 3 of \cite{Ambjorn:2001xj} and
Fig.\ 3 of \cite{Ambjorn:2000dj} should have $p$ instead of $p^2$
as the label.}
The VEV $\langle P (\vec{p})\rangle$
depends only on 
$p \defeq \sqrt{p_\mu p_\mu}$ due to the SO(4) 
invariance.
%\footnote{As explained later, 
%this statement is true even if the SO(4)
%symmetry is spontaneously broken.}
It starts at 1 for $p=0$ and 
drops down to zero at some value $\bar{p}$.
Then $1/\bar{p}$ is a measure for the extent of the space-time.
In Ref.\ \cite{AABHN}, it was shown that 
the one-point function $\langle P (\vec{p})\rangle$, as well as 
other Wilson loop correlators,
converge to a certain function of $p$ in the large $N$ limit,
if the scale parameter $g$ is taken to be proportional to $1/\sqrt{N}$.
%(i.e.\ $g^2 N$ is fixed).
This means in particular that 
$\bar{p} \propto (\sqrt{g} N^{1/4})^{-1}$. 
(If we set $g=(N/48)^{-1/2}$, as we do in Figures \ref{P-k-fig} and 
\ref{P-k-fig2}, we find $\bar p  \approx 0.7$ for 
$N=16,  \dots , 48$.)
Therefore, the large $N$ behavior of the space-time extent
probed by the Polyakov line is
$1/\bar{p} \propto \sqrt{g} N^{1/4}$, which is consistent with the result
obtained from $R_{\new}$ defined in eq.\ (\ref{Rnewdef}).
%prediction based on the low-energy effective theory.

Let us formulate the SSB of Lorentz symmetry
by using the Wilson loops as a probe.
For each configuration $A_\mu$ we perform a SO(4) transformation
$\tilde{A} _\mu = \Lambda _{\mu\nu} A_\nu$ so that
$\tilde{I} _{\mu \nu} = \frac{1}{N}\Tr (\tilde{A} _\mu \tilde{A} _\nu )$ 
becomes diagonal : $\tilde{I} = \mbox{diag}(\omega _ 1 , \omega _ 2 ,
\omega _ 3 , \omega _ 4)$, where 
$\omega _ 1 >  \omega _ 2 > \omega _ 3 > \omega _ 4 > 0$.
Then we define
\beq
\tilde{P} _\mu (p) = \frac{1}{N} \, \Tr \, \exp (i \, p \tilde{A} _\mu) \ .
\eeq
As a consequence of the diagonalization,
the VEV's $\langle \tilde{P} _\mu (p)\rangle$ ($\mu = 1 , \dots , 4$)
do depend on $\mu$ for finite $N$.
If these functions are different even in the large $N$ limit,
we may conclude that the SSB of SO(4) symmetry occurs.

This method is analogous to the Ising model,
where the magnetization $\langle |M| \rangle$
($M$ is the sum of spins divided by their number), 
serves as an order parameter for the SSB of $Z_2$ symmetry.
Taking the absolute value of $M$ corresponds to making an 
appropriate SO(4) transformation from $A_\mu$ to $\tilde{A}_\mu$.
Note, on the other hand, 
that $\langle M \rangle = 0$ for any finite lattice volume 
even if the SSB takes place (for infinite lattice volume).
Similarly, the VEV $\langle P (\vec{p})\rangle$ in eq.\ (\ref{Pol1}) 
depends only on $p =\sqrt{p_\mu p_\mu }$,
even if the SO(4) symmetry is spontaneously broken.

Our results are shown in Figure \ref{P-k-fig}.
(In this case we just present the results obtained at
$\Delta \tau = 0.002$, which appears to be sufficiently small.)
Here we set $g=(N/48)^{-1/2}$
and plot $\langle \tilde P_{\mu} (p) \rangle $
against $N$ at three different values of the momentum $p$, 
all of them below $\bar p$.
We observe no trend of SSB up to $N=48$.
To illustrate this observation in yet another way, we show
in Figure \ref{P-k-fig2} the functions 
$\langle \tilde P _\mu (p)\rangle$ 
for $N=16$ and for $N=48$. We see that the results for the 
four Polyakov lines move closer together as $N$ increases.

Finally, we clarify the relation of the above result
to the previous result (\ref{divlambda}).
% obtained in the previous Section.
Let us denote the eigenvalues of $\tilde{A}_\mu$ as 
$\alpha _{\mu i}$ ($i = 1,\cdots , N$) and introduce
the probability distribution of the $\alpha_{\mu i}$ as
\beq
f _\mu (x) = \left\langle \frac{1}{N} \sum _{i=1} ^N
\delta (x - \alpha _{\mu i}) \right \rangle \ .
\eeq
Then we can write $\langle \tilde{P} _\mu (p) \rangle$ as
\beq
\langle\tilde{P} _\mu (p) \rangle
= \left\langle \frac{1}{N} \sum_{i=1}^N \ee ^{i \, p \alpha _{\mu i}} 
\right\rangle
= \int _{-\infty} ^{\infty}  
\dd  x  \, f _\mu (x) \, \ee ^{i \, p x} \ .
\eeq
Thus the VEV of the Polyakov line is just the Fourier transform
of the distribution $f _\mu (x)$.
Our observation that $\langle \tilde{P} _\mu (p) \rangle$
for $\mu = 1,2,3,4$ approach the same function of $p$
at $N = \infty$ implies that their inverse Fourier transforms
$f _\mu (x)$ for $\mu = 1,2,3,4$ also approach to the same function
of $x$ at $N \rightarrow \infty$.
On the other hand, our result in the Figure \ref{T-EV} suggests that 
$f _1 (x) \sim C(N)/|x|^3$ at large $x$, but the $f_{\mu} (x)$  
with $\mu = 2,3,4$ have a sharper fall-off at large $x$.
This suggests that the coefficient $C(N)$ vanishes
in the large $N$ limit (at fixed $g^{2}N$).

We can rephrase the above statement in the following way.
Note that $ \langle \omega _ \mu \rangle =
 \langle \frac{1}{N} \Tr (\tilde{A} _\mu)^2 \rangle $
(here we do not sum over the index $\mu$ on the r.h.s.)
can be written as
\beq
\langle \omega _ \mu \rangle = \left\langle 
\frac{1}{N} \sum _{i=1} ^N (\alpha _{\mu i})^2 \right\rangle 
= \int _{-\infty} ^{\infty}  \dd  x  \, x^2 f _\mu (x)
= -  \left. \frac{\dd ^2}{\dd p ^2} \langle\tilde{P} _\mu (p) \rangle 
\right|_{p=0} \ .
\eeq
The logarithmic divergence of $\langle \omega _ 1 \rangle $
found in the previous Section (consider footnote \ref{AmuAnu})
implies that $\langle \tilde{P}_1 (p)\rangle$ has a non-analyticity
at $p=0$ of the form 
\beq
\langle \tilde{P}_1(p) \rangle \sim 1 + \tilde{C}(N) 
p^2 \ln p + \cdots \ .
% \qquad \qquad (\tilde{C}(N) = const.)
\label{Pol_expand}
\eeq
On the other hand, $\langle \tilde{P}_\mu (p)\rangle$ with
$\mu = 2,3,4$ do not have this non-analyticity,
since $\langle \omega _ 2 \rangle$,
$\langle \omega _ 3 \rangle$,
$\langle \omega _ 4 \rangle$
are finite.
Our observation that
the four Polyakov lines
$\langle \tilde{P} _\mu (p)\rangle$ ($\mu = 1 , \dots , 4$)
converge to a single function of $p$ suggests that
the coefficient $\tilde{C}(N)$ in eq.\ (\ref{Pol_expand})
vanishes in the large $N$ limit (at $g^{2}N$ fixed).

These statements can be checked by extracting 
the coefficient of the logarithmic divergence in eq.\ (\ref{divlambda}).
If we fit the data for $\langle \lambda _1 \rangle /g$ to the expected
asymptotic function \ $- c_{1}(N)\ln \Delta \tau + c_{0}(N)$, we find 
that the normalized coefficient $c_1(N)/\sqrt{N}$ amounts to
2.2(3), 2.52(7), 1.75(7) for $N=16$, 24 and 32, respectively.
Hence the behavior $C, \tilde{C} \rightarrow 0 $
as $N \rightarrow \infty$ is conceivable.

To summarize, our observation in this Section implies that the quantities
\beq
 \int _{-\infty} ^{\infty}  \dd  x  \, x^2 
\left[ \lim _{N\rightarrow \infty} f _\mu (x) \right]
= - \left. \frac{\dd ^2}{\dd p ^2} 
\left[ \lim _{N\rightarrow \infty}
\langle\tilde{P} _\mu (p) \rangle  \right]
\right|_{p=0} 
\mbox{~~~~~}(\mu = 1,\cdots,4)
\label{finiteomega_1}
\eeq
are all finite and equal.
Notice that 
in eq.\ (\ref{finiteomega_1}),
the limit $N\rightarrow \infty$ is taken {\em before} setting $p=0$.
The point is that one should first take the large $N$ limit
of the Wilson loop to make it actually physical, and {\em then}
its derivatives at $p=0$ inherit a physical meaning, too.
This does not need to be true for derivatives at $p=0$ for
finite $N$. In fact, our results suggest that the $p \rightarrow 0 $ limit
and the large $N$ limit do not commute.

\section{Discussion}
\label{summary}

In this paper, we wanted to clarify %the subtlety in 
the issue of spontaneous Lorentz symmetry breakdown in supersymmetric 
matrix models, which was raised by Refs.~\cite{Burda:2000mn}.
We propose a physical criterion for SSB,
which we consider as a solution to this problem.
In the particular case of the 4D SUSY model, 
configurations with only one-dimensional extent 
dominate when one adopts a conventional criterion for the SSB 
using the moment of inertia tensor,
as was suggested by Refs.~\cite{Burda:2000mn}.
However, contributions of those configurations to physical quantities 
such as Wilson loop correlators seem to be strongly suppressed
in the large $N$ limit. Indeed, if we rely on our physical criterion
using Wilson loops as a probe, we do not observe any trend of SSB;
the space-time probed by Wilson loops appears to be four-dimensional.

Let us comment on the cases $D=6$ and $D=10$. We recall that 
in the $D$-dimensional SUSY models in general, $D=4,6,10$,
the eigenvalue distribution of $A_\mu$ has a slow fall-off
with the power $-(2D-5)$ independent of $N$ \cite{Eigen}.
Therefore, as pointed out also in Ref.\ \cite{Burda:2000mn},
a problem may arise if one considers a tensor such as
$(I^n)_{\mu\nu}$, where 
$I_{\mu\nu}=\frac{1}{N} \Tr (A_{\mu} A_{\nu})$ and $n \ge D-3$.
The VEVs of the corresponding eigenvalues can be written as
\beq
\langle (\omega _ \mu)^n \rangle = \left\langle 
\frac{1}{N} \sum _{i=1} ^N (\alpha _{\mu i})^{2n} \right\rangle 
= \int _{-\infty} ^{\infty}  \dd  x  \, x^{2n} f _\mu (x)
= (-1)^{n}   \left. \frac{\dd ^{2n}}{\dd p ^{2n}} 
\langle\tilde{P} _\mu (p) \rangle 
\right|_{p=0} \ .
\eeq
However, one has to take the large $N$ limit before setting $p=0$,
in order to have a correct interpretation in string theory.
The possible divergence of $\langle (\omega _ 1)^n \rangle$
for finite $N$ does not immediately imply an appearance of 
the SSB of Lorentz symmetry in physical quantities,
as we have seen in $D=4$.
On the other hand, the conventional moment of inertia tensor
$I_{\mu\nu}$ (or $T_{\mu\nu}$ in Section \ref{ssb}) 
does not have such a problem
in $D=6,10$, since the $p\rightarrow 0 $ limit and the large $N$ limit 
commute.
If one observes an SSB with the conventional moment 
of inertia tensor, it immediately implies an SSB in physical quantities.

The absence of SSB in 4D SUSY model (with the physical criterion)
is consistent with the conjecture that the phase
of the determinant in the fermionic partition function plays
a crucial r\^{o}le in a possible SSB of Lorentz symmetry.
This conjecture is supported by the following results:
\begin{enumerate}
\item The bosonic model does not show SSB \cite{HNT}.
\item The SUSY model in $D=6$ and $D=10$ (using a low-energy
effective theory) does not exhibit SSB if one omits
the phase of the determinant \cite{branched}. 
\item SSB does appear in the $\nu$-deformed SUSY model
in $D=6$ and $D=10$ 
in the large $\nu$ limit, where $\nu$ couples to the phase 
of the fermionic partition function \cite{NV}.
\item No SSB occurs in the 4D SUSY model (where the fermionic
partition function is real positive), as discussed in the
present work, where we do not use any simplification of the
full model.
\end{enumerate}
When one integrates over the bosonic matrices in the
6D or 10D IIB matrix model,
the phase of the fermion integral fluctuates rapidly
except for the vicinity of configurations,
for which the phase becomes stationary. 
In fact, this happens for any $d$-dimensional configurations with
$3 \le d \le D-2$ \cite{NV}.
%\footnote{This effect
%has been demonstrated explicitly in a simplified model \cite{brane}.}
Those configurations are therefore considerably enhanced
compared to the case where the phase is omitted.
By using a saddle-point approximation,
it was found that only the dimensionalities
within the above range are possible.
Note in particular that $d=4$ is not excluded.

%\vspace*{7mm}

\section*{Acknowledgement}
We would like to thank 
P. Bialas, Z. Burda, B. Petersson, J. Tabaczek and J.F. Wheater
for useful discussions. The computation for generating configurations
has been carried out partly on 
VPP700E at The Institute of Physical and Chemical Research (RIKEN),
and on SX4 at the Research Center for Nuclear Physics (RCNP) of Osaka
University. K.N.A.'s research was partially supported by RTN grants
HPRN-CT-2000-00122 and HPRN-CT-2000-00131, the INTAS contract
N 99 0590, and a National Fellowship Foundation of Greece (IKY)
postdoctoral fellowship. 
%and EU grant HPRNCT-199900161. 
J.A.\ acknowledges support by the EU network on 
``Discrete Random Geometry'', grant HPRN-CT-1999-00161 
(which also supported KNA), 
by ESF network no.82 on ``Geometry and Disorder'' and 
by ``MaPhySto'', the Center of Mathematical Physics 
and Stochastics, financed by the National Danish Research Foundation.

%\vspace*{7mm}


\begin{thebibliography}{50}

\bibitem{BFSS} T.\ Banks, W.\ Fischler, S.H.\ Shenker and L.\ Susskind,
``M Theory as a Matrix Model: A Conjecture'',
Phys.\ Rev.\ {\bf D55} (1997) 5112,
{\tt hep-th/9610043}.

\bibitem{IKKT} N.\ Ishibashi, H.\ Kawai, Y.\ Kitazawa and A.\ Tsuchiya,
``A Large-$N$ Reduced Model as Superstring'',
Nucl.\ Phys.\ {\bf B498} (1997) 467, {\tt hep-th/9612115}.

\bibitem{SFT}
J.\ Greensite and F.R.\ Klinkhamer,
``Superstring Amplitudes and Contact Interactions'', 
Nucl.\ Phys. {\bf B304} (1988) 108.\\
V.A.\ Kosteleck\'{y} and S.\ Samuel, 
``Spontaneous Breaking of Lorentz Symmetry in
String Theory'',  
Phys.\ Rev. {\bf D39} (1989) 683.

\bibitem{YMd} G.K. Savvidy, ``Yang-Mills Quantum Mechanics'',
Phys. Lett. {\bf B159} (1985) 325.\\
M. Claudson and M.B. Halpern,
``Supersymmetric ground state wave functions'',
Nucl. Phys. {\bf B250} (1985) 689.

\bibitem{NV}
J.\ Nishimura and G.\ Vernizzi,
``Spontaneous Breakdown of Lorentz Invariance in IIB Matrix Model'',
%JHEP {\bf 0004} (2000) 015.
J. High Energy Phys. {\bf 04} (2000) 015,
{\tt hep-th/0003223}. 

\bibitem{brane} J.\ Nishimura and G.\ Vernizzi,
``Brane World Generated Dynamically from String Type IIB Matrices'',
Phys. Rev. Lett. {\bf 85} (2000) 4664,
{\tt hep-th/0007022}.

\bibitem{RS}
L.\ Randall and R.\ Sundrum,
``An Alternative to Compactification'',
Phys.\ Rev.\ Lett. {\bf 83} (1999) 4690,
{\tt hep-th/9906064}.

\bibitem{IIKK}
N.\ Ishibashi, S.\ Iso, H.\ Kawai and Y.\ Kitazawa,
``String Scale in Noncommutative Yang-Mills'',
Nucl.\ Phys.\ B {\bf 583} (2000) 159.
{\tt hep-th/0004038}.

\bibitem{HNT} T.\ Hotta, J.\ Nishimura and A.\ Tsuchiya, 
``Dynamical Aspects of Large $N$ Reduced Models'',
Nucl.\ Phys.\ {\bf B545} (1999) 543,
{\tt hep-th/9811220}.

\bibitem{bos} W.\ Krauth and M.\ Staudacher,
``Finite Yang-Mills Integrals,''
Phys.\ Lett.\ {\bf B435} (1998) 350,
 {\tt hep-th/9804199};
``Statistical Physics Approach to M-theory Integrals'',
cond-mat/0010127.\\
S.\ Horata and H.\ Egawa, 
``Numerical Analysis of the Double Scaling Limit
in the IIB Matrix Model'', {\tt hep-th/0005157}.\\
S.\ Oda and F.\ Sugino,
``Gaussian and Mean Field Approximations for Reduced
Yang-Mills Integrals'', 
J. High Energy Phys. {\bf 0103} (2001) 026, 
{\tt hep-th/0011175}.\\
%\bibitem{Anagnostopoulos:2000mn}
K.N.~Anagnostopoulos, J.~Nishimura and P.~Olesen,
``Noncommutative String Worldsheets from Matrix Models,''
{\tt hep-th/0012061}, to appear in J. High Energy Phys.\\
P.\ Austing and J.F.\ Wheater,
``The Convergence of Yang-Mills Integrals'',
J. High Energy Phys. {\bf 0102} (2001) 028, 
{\tt hep-th/0101071}.
%%CITATION = HEP-TH 0012061;%%

\bibitem{AW} P.\ Austing and J.F.\ Wheater,
``Convergent Yang-Mills Matrix Theories'', {\tt hep-th/0103159}.

\bibitem{branched}
J.\ Ambj\o rn, K.N.\ Anagnostopoulos, W.\ Bietenholz,
T.\ Hotta and J.\ Nishimura,
``Monte Carlo studies of the IIB matrix model at large $N$'',
J. High Energy Phys. {\bf 0007}, 011 (2000),
{\tt hep-th/0005147}.

\bibitem{AABHN} J.\ Ambj\o rn, K.N.\ Anagnostopoulos, W.\ Bietenholz,
T.\ Hotta and J.\ Nishimura,
``Large $N$ Dynamics of Dimensionally Reduced 4D SU($N$) Super Yang-Mills 
Theory'', J. High Energy Phys. {\bf 0007}, 013 (2000),
{\tt hep-th/0003208}.

\bibitem{Burda:2000mn} Z.\ Burda, B.\ Petersson and J.\ Tabaczek,
``Geometry of Reduced Supersymmetric 4D Yang-Mills Integrals'',
{\tt hep-lat/0012001}. {\it See also} \\
P. Bialas, Z. Burda, B. Petersson and J. Tabaczek,
``Large $N$ Limit of the IKKT Model'',
Nucl. Phys. {\bf B592} (2001) 391, {\tt hep-lat/0007013}.

\bibitem{Eigen} W.\ Krauth and M.\ Staudacher,
``Eigenvalue Distributions in Yang-Mills Integrals'',
Phys.\ Lett.\ {\bf B453} (1999) 253, 
{\tt hep-th/9802113}.

\bibitem{FKKT} M.\ Fukuma, H.\ Kawai, Y.\ Kitazawa and A.\ Tsuchiya,
``String Field Theory from IIB Matrix Model'',
Nucl. Phys. {\bf B510} (1998) 158, 
{\tt hep-th/9705128}.

\bibitem{KNS} W.\ Krauth, H.\ Nicolai and M.\ Staudacher,
``Monte Carlo Approach to M-Theory'',
Phys.\ Lett.\ {\bf B431} (1998) 31,
{\tt hep-th/9803117}.

\bibitem{AIKKT} H.\ Aoki, S.\ Iso, H.\ Kawai, Y.\ Kitazawa and T.\ Tada,
``Space-time Structures from IIB Matrix Model'',
Prog.\ Theor.\ Phys.\ {\bf 99} (1998) 713,
{\tt hep-th/9802085}.

\bibitem{Aoki:1999bq}
H.~Aoki, S.~Iso, H.~Kawai, Y.~Kitazawa, A.~Tsuchiya and T.~Tada,
``IIB Matrix Model'',
Prog.\ Theor.\ Phys.\ Suppl.\ {\bf 134} (1999) 47,
{\tt hep-th/9908038}.
%%CITATION = HEP-TH 9908038;%%

\bibitem{KU} N.\ Kitsunezaki and S.\ Uehara,
``Large $N$ Structure of IIB Matrix Model'',
{\tt hep-th/0010038}.

\bibitem{AV} J.\ Ambj\o rn and S.\ Varsted, 
``Dynamical triangulated fermionic surfaces'',
Phys.\ Lett.\ {\bf B257} (1991) 305.

\bibitem{hybridR}
S.\ Gottlieb, W.\ Liu, D.\ Toussaint, R.L.\ Renken
and R.L.\ Sugar, ``Hybrid Molecular Dynamics Algorithms
for the Numerical Simulation of Quantum Chromodynamics'',
Phys.\ Rev.\ {\bf D35} (1987) 2531.

\bibitem{Iso:2000xs}
S.\ Iso and H.\ Kawai,
``Space-time and Matter in IIB Matrix Model: 
Gauge Symmetry and  Diffeomorphism'',
Int.\ J.\ Mod.\ Phys.\ {\bf A15} (2000) 651,
{\tt hep-th/9903217}.

\bibitem{Ambjorn:2001xj}
J.~Ambj\o rn, K.N.~Anagnostopoulos, W.~Bietenholz, T.~Hotta 
and J.~Nishimura,
``Simulating Simplified Versions of the IKKT Matrix Model'',
Nucl.\ Phys.\ (Proc.\ Suppl.) {\bf 94} (2001) 685,
{\tt hep-lat/0009030}.

\bibitem{Ambjorn:2000dj}
J.~Ambj\o rn, K.N.~Anagnostopoulos, W.~Bietenholz, T.~Hotta 
and J.~Nishimura,
``Monte Carlo Studies of the Dimensionally Reduced 4D SU(N) 
Super  Yang-Mills Theory'',
{\tt hep-th/0101084}.

% \bibitem{2DEK}
% T.\ Nakajima and J.\ Nishimura,
% ``Numerical Study of the Double Scaling Limit in Two-Dimensional
% Large $N$ Reduced Model'',
% Nucl.\ Phys.\ {\bf B528} (1998) 355, {\tt hep-th/9802082}.

% \bibitem{KS} W.\ Krauth and M.\ Staudacher,
% ``Finite Yang-Mills integrals,''
% Phys.\ Lett.\ {\bf B435} (1998) 350, {\tt hep-th/9804199}.

% \bibitem{analytic}
% J.\ Hoppe, V.A.\ Kazakov and I.K.\ Kostov, 
% ``Dimensionally Reduced SYM(4) as Solvable Matrix Quantum Mechanics'',
% Nucl.\ Phys.\ {\bf B571} (2000) 479,
% {\tt hep-th/9907058}.

\end{thebibliography}
\end{document}